\documentclass[sigplan,screen,nonacm]{acmart}
\UseMicrotypeSet[protrusion]{basictext}

\newif\ifrevisionnotes
\makeatletter
\if@ACM@anonymous \revisionnotestrue \fi
\makeatother
\ifrevisionnotes
  \newcommand{\revlabel}[1]{\footnotesize\sffamily\color{blue!60!black}#1}
  \newcommand{\revpoint}[1]{\marginpar[\raggedleft\revlabel{#1}]%
    {\raggedright\revlabel{#1}}}
  \newcommand{\revised}[1]{\textcolor{blue!60!black}{#1}}
\else
  \newcommand{\revpoint}[1]{}
  \newcommand{\revised}[1]{#1}
\fi

\usepackage{xurl}

\usepackage{tikz}
\usepackage{stfloats}
\usepackage{booktabs}
\usepackage{makecell}
\usepackage{multirow}
\usepackage{subcaption}
\usepackage{xspace}
\usepackage{listings}
\usepackage{algorithm}
\usepackage{algpseudocode}
\usepackage{wasysym}
\usepackage{pgfplots}
\usepackage{siunitx}
\pgfplotsset{compat=1.18}

\usetikzlibrary{patterns,arrows.meta}

\makeatletter
\if@ACM@anonymous
  \newcommand{\mold}{pld\xspace}
  \newcommand{\moldversion}{(the revision included in the submitted artifact)\xspace}
\else
  \newcommand{\mold}{mold\xspace}
  \newcommand{\moldversion}{2.42.0\xspace}
\fi
\makeatother

\newcommand{\lld}{lld\xspace}
\newcommand{\gold}{gold\xspace}
\newcommand{\gnuld}{GNU~ld\xspace}
\newcommand{\code}[1]{\texttt{#1}}

\DeclareRobustCommand{\pfull}{\tikz[baseline=-0.55ex]\draw[fill=black] (0,0) circle (0.75ex);}
\DeclareRobustCommand{\phalf}{\tikz[baseline=-0.55ex]{\draw (0,0) circle (0.75ex);
  \fill (0,0) -- (90:0.75ex) arc (90:270:0.75ex) -- cycle;}}
\DeclareRobustCommand{\pnone}{\tikz[baseline=-0.55ex]\draw (0,0) circle (0.75ex);}

\begin{document}

\title{\mold: A Massively Parallel Linker}

\author{Rui Ueyama}
\orcid{0009-0004-3216-9537}
\affiliation{%
  \institution{The University of Tokyo}
  \city{Tokyo}
  \country{Japan}}
\email{ruiu@is.s.u-tokyo.ac.jp}

\begin{abstract}
Linking is a critical step in the software build process that combines
compiled object files into a single executable or shared library. Despite
decades of engineering effort, link times remain a significant
bottleneck in the edit-compile-debug cycle, particularly for large C++
programs. Existing linkers exploit limited parallelism,
leaving most CPU cores idle during linking. We present \mold, a
Unix/Linux linker that applies data parallelism systematically
across the entire linking pipeline. We first analyze the architectural constraints that prevent
existing linkers from scaling, including entangled symbol resolution
and archive processing, and then show how a clean-slate design that decouples them
overcomes these limitations. On large
real-world programs, \mold links multi-gigabyte debug binaries in at most
a few seconds, and often in under a second. It is 2.4--16.1x
faster than the
state-of-the-art \lld linker, and up to 112x faster than the
traditional \gnuld.
An ablation study shows that no single optimization
dominates; the speedup comes from the cumulative effect of
parallelizing all passes.
\end{abstract}



\maketitle

\ifrevisionnotes
{\small\color{blue!60!black}\noindent
Passages that directly address individual items in the revision plan
are shown in blue and labeled in the margin. All performance
experiments were rerun for this revision using the most recent version
of \mold, which incorporates performance improvements over the version
evaluated in the original submission. The workload suite and
comparison methodology were also revised to address reviewer concerns,
so the new measurements supersede those in the original submission and
are not directly comparable to them. The manuscript contains broader
substantive revisions that are not marked individually.
\par\medskip}
\fi

\section{Introduction}
\label{sec:intro}

Building large software systems from source code is a
multi-stage process that typically involves compiling individual source
files into object files and then linking them together
into an executable or shared library. Despite extensive research on
compilation, the linking step has remained largely neglected. Yet for
large programs, linking alone can take tens of seconds, becoming the
bottleneck in the edit-compile-debug cycle that developers repeat
dozens of times a day. A debug build of
TensorFlow, for example, comprises 24\,GiB of object files, which
the linker combines into a single 9.9\,GiB shared library.
Linking such a program with the state-of-the-art LLVM
\lld~\cite{ueyama2017lld} takes 52 seconds on a 64-core
machine, leaving most of its cores idle.

This paper presents \mold, a production-quality Unix/Linux linker
designed from scratch around a single principle: structure every major
pass as a data-parallel loop over homogeneous arrays of elements
(symbols, relocations, sections), using concurrent data structures and
atomic operations for the few points requiring synchronization. The
key architectural decision is to decouple parsing from symbol
resolution: \mold eagerly parses all input files---including every
member of every archive---in parallel, and only then performs symbol
resolution as a separate parallel pass using atomic compare-and-swap.
This eliminates the entangled processing that forces sequential
execution in prior linkers. The same data-parallel structure is
applied to every subsequent pass: relocation scanning, section garbage
collection, identical code folding, string merging, layout
computation, and output generation. By Amdahl's
law~\cite{amdahl1967}, parallelizing only a subset of passes leaves
the serial remainder as a hard ceiling on speedup; \mold avoids this
by parallelizing all of them.

The contributions of this paper are:
\begin{itemize}
\item We analyze the architectural constraints that prevent existing
  linkers from exploiting modern multi-core hardware
  (\S\ref{sec:why-not-parallel}).
\item We present a linker architecture that decouples traditionally
  entangled phases and applies data parallelism to every major pass,
  and describe the concurrent algorithms used for each
  (\S\ref{sec:design}, \S\ref{sec:impl}).
\item We provide a comprehensive evaluation on nine large real-world
  workloads, including an ablation study quantifying each technique's
  contribution, a scalability analysis from 1 to 64 threads, and a
  head-to-head per-pass comparison with \lld (\S\ref{sec:eval}).
\end{itemize}

\section{Background}
\label{sec:background}

This section outlines the standard linking pipeline, reviews
existing linkers and why they are not more parallel, and then
introduces key ELF concepts referenced by later sections.

\subsection{The Linking Pipeline}

A linker reads object files and shared libraries, resolves symbol
references, applies relocations, and produces an executable or shared
library. The major phases are:

\begin{enumerate}
\item \textbf{Input parsing}: Read and parse all input object files,
  shared libraries, and archives. For each file, extract the symbol
  table and section table.

\item \textbf{Symbol resolution}: Build a global symbol table that
  maps each symbol name to its defining file. Deduplicate COMDAT groups,
  keeping one copy of each. Resolve references from archives
  on demand.

\item \textbf{Section processing}: Optionally perform garbage
  collection and identical code folding. Merge string
  sections, deduplicating identical string literals across input
  files.

\item \textbf{Layout computation}: Construct synthetic sections (GOT,
  PLT, dynamic symbol tables), determine the output file layout by
  computing sizes and offsets of all output sections, and assign virtual
  addresses.

\item \textbf{Output generation}: Copy section contents from input
  files to the output file, applying relocations to fill in symbol
  addresses. Write synthetic section contents and
  file headers.
\end{enumerate}

\subsection{Existing Linkers}

\textbf{GNU ld} (also known as BFD ld) is the traditional Unix
linker, distributed as part of the GNU Binutils package. It is
single-threaded and processes input files sequentially, resulting in
poor performance on modern hardware.

\textbf{GNU gold}~\cite{taylor2008gold} was written from scratch at
Google in 2008 as a faster alternative to \gnuld. As an ELF-only
linker designed with speed in mind, it achieved significant speedups
over \gnuld. \gold also introduced multi-threaded support, though it parallelizes
only a limited number of passes. After Google
migrated to \lld, \gold lost its primary maintainer and was marked
deprecated in 2025~\cite{binutils244}.

\textbf{LLVM lld}~\cite{ueyama2017lld} is the linker from the LLVM
project. It parallelizes several passes including some post-parse
input-file work, relocation scanning, and output writing.
It achieves 2--9x speedup
over \gold on large programs. However, key passes such as symbol
resolution remain sequential~\cite{maskray2021lld}.

A summary of which passes are parallelized in each linker is shown
in Table~\ref{tab:parallel-comparison}.

\begin{table}[t]
\centering
\caption{Parallelization of major linker passes in the evaluated
  versions (\gnuld/\gold 2.46.1, \lld 22.1.8, and \mold \moldversion).
  \pfull~= parallelized, \phalf~= partially or conditionally
  parallelized, \pnone~= sequential, ---~= unsupported.}
\label{tab:parallel-comparison}
\small
\begin{tabular}{@{}lcccc@{}}
\toprule
Pass & \gnuld & \gold & \lld & \mold \\
\midrule
Input file parsing     & \pnone & \pfull & \phalf & \pfull \\
Symbol resolution      & \pnone & \pnone & \pnone & \pfull \\
Relocation scanning    & \pnone & \pnone & \pfull & \pfull \\
Section GC             & \pnone & \pnone & \pnone & \pfull \\
Range extension thunks & \pnone & \pnone & \pnone & \pfull \\
Identical code folding & ---    & \pnone & \pfull & \pfull \\
String merging         & \pnone & \pnone & \pfull & \pfull \\
Section copy + reloc.\ apply & \pnone & \pfull & \pfull & \pfull \\
Build-ID computation   & \pnone & \pfull & \pfull & \pfull \\
\bottomrule
\end{tabular}
\end{table}

\subsection{Why Existing Linkers Are Not More Parallel}
\label{sec:why-not-parallel}

Given that data parallelism is conceptually straightforward, a natural
question is why existing linkers do not already exploit it fully.
We identify three reasons.

\paragraph{Order-sensitive semantics}
Traditional Unix linking is defined in terms of a single
left-to-right pass over the command line. An archive
satisfies only the references that are still undefined when the
scan reaches it, and an extracted member may itself introduce new
undefined references, so member extraction follows use-def chains
across archives that a sequential scan naturally resolves. A
single scan cannot handle circularly-dependent archives, so users
must wrap them in \code{-start-group} and \code{-end-group},
which request that the group be rescanned until no new undefined
references are introduced. Reproducing the exact set of archive members
extracted by this process is hard with a parallel algorithm in some
corner cases.

\paragraph{No formal specification}
Unlike programming languages, whose semantics are typically
defined by a formal standard, the ELF specification~\cite{elfspec}
defines only the object file format, not how a linker should process
its inputs.
Kell et al.~\cite{kell2016missinglink} give a formal semantics for ELF
linking, but their model covers only static linking of small C
programs and aims to formalize existing practice rather than define a
normative standard.
In practice, \gnuld is widely considered
the de facto reference implementation.

\revpoint{B: sem.}\revised{This absence of a specification makes established conventions hard to
change, because it is not clear what is allowed to change.
Parallelizing the linker fully can alter corner-case outcomes, such as
which definition prevails when two archives wrapped with \code{-start-group}
and \code{-end-group} define the same symbol.
In practice, the only way to evaluate such a deviation is to deploy
the linker and see whether anything breaks. Users treat any observable difference as a linker bug, so
for an established linker, the experiment is rarely worth the risk.
A new linker's early adopters knowingly accept such differences, which
is one reason why drastic improvements tend to come from new linkers
rather than existing ones.}

\paragraph{Pervasive parallelism is hard to retrofit}
\gold and \lld illustrate the limit of incremental parallelization.
Both aim for speed and have received years of optimization.
They use threads in passes that can be isolated, yet symbol resolution
and other central passes remain sequential, as
Table~\ref{tab:parallel-comparison} shows. Parallelizing those passes
changes the order in which global decisions are made, the
representations that carry their results, and the interfaces to every
later pass. Making these changes amounts to rewriting much of the linker all
at once. At that scale, writing
a new linker from scratch can be easier than completing the retrofit.

\subsection{Key Concepts}
\label{sec:concepts}

Before describing our design, we briefly review the ELF and linking
concepts that the remaining sections build on.

\paragraph{Sections}
An ELF object file is divided into named regions called
\emph{sections}, containing code (\code{.text}), initialized data
(\code{.data}), uninitialized data (\code{.bss}), read-only data
(\code{.rodata}), and various metadata.

\paragraph{Symbols}
Symbols are named entities (functions, global variables) that may be
\emph{defined} (providing the content) or \emph{undefined} (referencing a
definition in another object file).

\paragraph{Relocations}
Relocations are metadata instructing the linker to patch specific
locations in section data, typically to fill in addresses of symbols
whose final locations are not known at compile time.

\paragraph{Archives}
An \emph{archive} (or static library, with the \code{.a} file
extension) is a bundle of object files. Unlike object files given
directly on the command line, object files in an archive are included
in the output only if they define a symbol referenced by another
included file. Archives allow programs to link against a large library
(such as libc.a) without pulling in every function it contains.

\paragraph{Position-independent code, GOT, and PLT}
Each process has its own virtual address space, so an executable can
be linked to a fixed address and use absolute addresses in its code.
Such code is called \emph{position-dependent}. Shared libraries,
however, cannot use fixed addresses because multiple libraries loaded
into the same process might collide. They are instead compiled as
\emph{position-independent code} (PIC) and loaded at an arbitrary
base address chosen at runtime. A PIC binary is
always loaded as a whole, so relative addresses within it are
invariant and can be embedded directly in the binary. However, absolute
addresses and references to symbols in other binaries are not known
until load time. Today, even executables are often built as
position-independent (PIE) to support address space layout
randomization (ASLR).

Generating position-independent code requires compiler cooperation
because the compiler must avoid emitting code that depends on absolute addresses.
When invoked with \code{-fPIC} or \code{-fPIE}, the compiler
generates position-independent code that uses indirection for
references whose targets may not be known at link time.
Accesses to such global variables go through the \emph{Global Offset
Table} (GOT), a per-binary table of pointers filled in at
runtime by the dynamic linker. Calls to such functions go through
the \emph{Procedure Linkage Table} (PLT), a set of stubs that jump
to addresses resolved at runtime. Both tables are accessed via
PC-relative addressing, so they work regardless of where the binary is
loaded. The static linker is responsible for scanning all relocations
to determine which symbols need GOT or PLT entries and generating
these tables accordingly.

\paragraph{COMDAT groups}
C++ compilers often emit multiple copies of the same function or data.
For instance, when a header file defining a template function is
included by multiple source files, each resulting object file contains
its own copy of every instantiation of that template. These duplicate definitions are placed
in \emph{COMDAT groups}: sets of sections tagged with a group name so
that the linker can keep one copy and discard the rest. In large C++
programs, COMDAT deduplication is a significant part of the linker's
work; a Firefox debug build contains 2.7 million COMDAT groups
across its input files, of which 1.6 million are redundant copies
that the linker discards.

\paragraph{Fine-grained sections}
When compilers are invoked with \code{-ffunction-sections} and
\code{-fdata-sections}, each function and global variable is placed
into its own section rather than being grouped into a single large
\code{.text} or \code{.data} section. This enables fine-grained
garbage collection (discarding individual unreferenced functions) and
identical code folding (merging functions with identical machine code)
by the linker.

\paragraph{Range extension thunks}
On typical RISC architectures, instructions are at most 4 bytes
wide, which leaves a call instruction only a modest displacement
field to encode the distance to its target. ARM64's \code{BL}
(branch-with-link) instruction, for example, has a 26-bit offset,
so a direct call can reach only targets within $\pm$128\,MiB of
the call instruction. Whether a call's target lies within that
range is not known until the linker lays out the output file.

The linker therefore checks each call relocation during layout.
When a target is out of reach, it inserts a
\emph{range extension thunk} (also called a \emph{veneer} or
\emph{trampoline})~\cite{aaelf64,song2026branches}
near the call site and redirects the original call to the thunk.
Each thunk is a short code sequence that
loads the full target address into a register and performs an indirect
jump, as in this ARM64 example:
\begin{lstlisting}
  caller: bl  thunk           // was 'bl callee'
  ...
  thunk:  adrp x16, callee
          add  x16, x16, :lo12:callee
          br   x16            // indirect jump
\end{lstlisting}
Among \mold's supported targets, ARM 32/64 and PowerPC 32/64 require
range extension thunks.

\section{Design Principles}
\label{sec:design}

A linker's workload is dominated by operations on large, homogeneous
arrays of data: tens of millions of relocations, millions of symbols,
hundreds of thousands of sections. Each element can typically be
processed independently or with only lightweight synchronization. This
structure is ideally suited to \emph{data parallelism}: applying the
same operation to each element of a large array in parallel.

The design of \mold is built around this data parallelism. Rather than trying to
overlap different phases (task parallelism), \mold executes passes
\emph{serially} but parallelizes \emph{within} each pass using
parallel-for loops. This serial-pass structure is easy to reason
about: at any point in execution, exactly one pass is running, and its
invariants are easy to state. \gold took the opposite approach,
employing task parallelism through a work queue with dependency
tokens~\cite{taylor2008gold}. In principle, independent tasks (e.g.,
reading different input files) can run concurrently. In practice,
however, most tasks are serialized by their dependency chains, and
benchmarks show negligible speedup or even minor slowdowns from
multi-threading. \gold disables multi-threading by default for this reason.
The lesson is that the
real opportunity lies not in overlapping coarse-grained phases but in
processing millions of homogeneous elements within a single phase.

\section{Parallelizing the Linking Pipeline}
\label{sec:impl}

We now describe how each major linker pass is parallelized in
\mold.

\subsection{Input File Parsing}
\label{sec:impl-parse}

To allow symbol resolution to run as a single parallel pass
(\S\ref{sec:symres}), \mold fully parses all input files, including archive
members, upfront and in parallel.

Parsing reads the ELF header, section headers, and symbol tables;
relocation tables are located but not processed until later
passes. Symbol names are \emph{interned} so that files that define
or reference the same name share one symbol object, reducing
symbol identity to pointer equality. These pointers form a
parallel array alongside each file's ELF symbol table, and all
later passes iterate over the interned symbols rather than the
raw ELF entries. Interning itself is parallel:
names are binned by hash into thread-local buffers during parsing,
and each bin is deduplicated independently afterwards, with no
shared concurrent data structure.

\subsection{Symbol Resolution}
\label{sec:symres}

Symbol resolution determines, for each symbol, which definition
prevails. Each file tries to install itself
as the \emph{owner} of the symbols it defines, using an atomic
compare-and-swap on the owner field of the shared symbol
object.
When multiple files define the same symbol, the precedence rules
in Table~\ref{tab:sym-rank} decide the winner.

\begin{table}[t]
\centering
\caption{Symbol precedence rules for parallel symbol resolution, from highest to lowest.
  Ties within the same rank are broken in favor of the definition
  whose input file appears earlier on the command line.}
\label{tab:sym-rank}
\small
\begin{tabular}{@{}cl@{}}
\toprule
Rank & Symbol type \\
\midrule
1 & Strong defined symbol \\
2 & Weak defined symbol \\
3 & Strong defined symbol in a shared library or archive \\
4 & Weak defined symbol in a shared library or archive \\
5 & Common symbol \\
6 & Common symbol in an archive \\
7 & Undefined (no definition seen) \\
\bottomrule
\end{tabular}
\end{table}

The ELF specification defines basic precedence rules (e.g., strong
definitions override weak ones) but leaves many interactions among
regular object files, archive members, and shared libraries
unspecified.
The rules in Table~\ref{tab:sym-rank} are therefore heuristic. We
tried several orderings and kept the one that disagrees least with
traditional linkers in practice. We validated this choice by
building the entire Gentoo repository, where only two of more than
19{,}000 packages failed to build because of this difference in
resolution semantics (\S\ref{sec:eval-compat}).

After symbol resolution, archive member inclusion is determined by
a liveness walk. Because the preceding compare-and-swap phase has
already set each symbol's owner to the file with the strongest
definition, every defined symbol now holds a direct pointer to its
owning file. The walk is therefore straightforward: non-archive
input files are unconditionally marked live, and when a file
becomes live, the linker follows the owner pointers of the
symbols it references to mark the defining files live in turn. This
makes the linker largely insensitive to the order of \code{-l}
flags: unlike a left-to-right scan of the command line, the walk
can reach an archive member that appears earlier on the
command line than the file that references it, and it follows
circular dependences among archives without any rescanning.
\mold accepts \code{-start-group} and \code{-end-group} for
compatibility, but they have no effect.

COMDAT deduplication works similarly. Group signatures are
interned like symbol names, so files
that define the same group share one group object. Live files
then claim each group with an atomic compare-and-swap, and files
that lose discard their copies of the group's member sections.

\subsection{String Merging}
\label{sec:strmerge}

ELF object files may contain \emph{mergeable string sections},
which hold string constants that the linker should deduplicate across
object files.
In a Firefox debug build, there are roughly 21 million such
strings, of which nearly three quarters are duplicates; merging them
can account for a significant fraction of the end-to-end link time.

\mold deduplicates strings by inserting them into a concurrent hash
map in parallel across all input files. We found that the general-purpose concurrent hash map in
oneAPI Threading Building Blocks (oneTBB)~\cite{onetbb} was not fast enough for this
workload, so we wrote a specialized one ourselves. The main optimization is to
avoid hash table resizing: we first estimate the number of distinct strings using
the HyperLogLog algorithm~\cite{flajolet2007hyperloglog} and allocate a table large enough from the
start. All input files then insert their strings in parallel using
atomic compare-and-swap; identical strings are mapped to the same
entry.

\subsection{Relocation Scanning}
\label{sec:relscan}

The linker must scan relocations to determine which symbols
require entries in the GOT, PLT, or other synthetic sections.
Whether an entry is required depends on the outcome of symbol
resolution, not on the relocation type alone. For example, a
relocation of type \code{R\_X86\_64\_PLT32} permits a PLT entry
rather than demands one: if the symbol was resolved to a definition
in the output executable itself, the call is bound directly and no
PLT entry is needed. For the Firefox debug build, this pass examines
11 million relocations.

Relocation scanning in \mold is a parallel-for loop over the
input sections. For each section, each relocation is examined, and if the
referenced symbol requires a GOT or PLT entry, an atomic flag on
the symbol object is set. Since each symbol's flag is set
monotonically (from ``no entry needed'' to ``entry needed''),
relaxed atomic bitwise-OR operations suffice; no locks are required.

After the parallel scan, the symbols with set flags are collected to
construct the GOT and PLT. This construction step is fast because the
number of symbols requiring GOT/PLT entry is typically orders of
magnitude smaller than the total number of symbols.

\subsection{Section Garbage Collection}

When programs are compiled with \code{-ffunction-sections} and
\code{-fdata-sections}, the linker can perform garbage
collection (GC) to discard unreferenced sections. This is modeled as
a mark-sweep traversal of a directed graph where sections are vertices
and relocations are edges.

\mold parallelizes the mark phase using a parallel-for-each loop
with a feeder pattern. The algorithm
starts from a set of root sections (those reachable from the entry
point, exported symbols, and sections with special semantics like
\code{.init\_array}) and performs a parallel reachability traversal.
When a thread marks a section as reachable, it feeds the section's
relocation targets into the work queue so they are visited in turn.

\subsection{Identical Code Folding}
\label{sec:icf}

Identical Code Folding (ICF) is an optional size optimization that
merges read-only sections with identical contents and relocations. The
challenge is that section identity is \emph{recursive}: for two
sections to be identical, their relocation targets must themselves be
identical. This is equivalent to computing bisimulation
equivalence over a directed graph in which sections are vertices, relocations are
edges, section contents are vertex colors, and relocation types are
edge labels.

\mold computes this equivalence with the hash-based refinement
algorithm that Sch\"atzle et al.~\cite{schaetzle2013bisim} used
for large-scale bisimulation reduction with MapReduce. The
algorithm is a form of \emph{color refinement}, also known as the
one-dimensional Weisfeiler-Leman algorithm~\cite{weisfeiler1968}.
It works as follows. On iteration $N$, each section's
cryptographic hash summarizes all walks of length up to $N$
originating from it. The first iteration hashes the section's
contents, flags, and relocation types, but not their targets.
Each subsequent iteration computes a new hash from the section's
previous hash and the previous hashes of its relocation targets,
extending the walk length by one. The hash sequence is monotonically
refining: each iteration can only distinguish sections that were
considered equal in the previous iteration, never merge previously
distinct ones. The process therefore converges when the number of
distinct hashes stops increasing, and sections with the same hash at
that point are considered identical. Cryptographic hashing makes
erroneous merges due to hash collisions negligibly unlikely.

Unlike classical partition refinement algorithms such as
Hopcroft's~\cite{hopcroft1971}, which explicitly split equivalence
classes, hashing avoids pairwise comparison, and every section
computes its new hash from the previous iteration's values alone.
Each iteration is embarrassingly parallel and needs no synchronization.

\begin{figure*}[t]
\centering
\begin{tikzpicture}[
  x=1cm, y=1cm,
  font=\small,
  cursor/.style={dash pattern=on 2.5pt off 2pt, thick},
  thunk/.style={fill=black},
  meas/.style={{Latex[length=2mm]}-{Latex[length=2mm]}, semithick},
  lbl/.style={font=\small},
]

\def\barbot{0}
\def\bartop{0.72}
\def\dx{1.9}          
\def\reach{6.45}       
\def\tw{0.22}         
\def\tcur{7.35}        
\pgfmathsetmacro\batchend{\tcur+\dx}        
\pgfmathsetmacro\reachlim{\batchend-\reach} 
\pgfmathsetmacro\lcur{\tcur+\reach}         

\fill[black!12] (0,\barbot) rectangle (\tcur,\bartop);
\fill[pattern=north east lines, pattern color=black!60]
     (\tcur,\barbot) rectangle (\batchend,\bartop);

\foreach \i in {1,...,5}
  \fill[thunk] ({\lcur-\i*\dx},\barbot) rectangle ({\lcur-\i*\dx+\tw},\bartop);
\foreach \i in {6,7} {
  \fill[white] ({\lcur-\i*\dx},\barbot) rectangle ({\lcur-\i*\dx+\tw},\bartop);
  \draw[thin]  ({\lcur-\i*\dx},\barbot) rectangle ({\lcur-\i*\dx+\tw},\bartop);
}
\fill[thunk] (\lcur,\barbot) rectangle (\lcur+\tw,\bartop);

\draw[thick] (0,\bartop) -- (\lcur+\tw,\bartop) -- (\lcur+\tw,\barbot) -- (0,\barbot);

\foreach \x in {14.4,15.2,16.0,16.8}
  \draw[dash pattern=on 2pt off 1.6pt, semithick]
    (\x,\barbot) rectangle (\x+0.58,\bartop);

\draw[cursor] (\reachlim,-1.12) -- (\reachlim,\bartop+0.95);
\draw[cursor] (\tcur,-1.12) -- (\tcur,\bartop+0.95);
\draw[cursor] (\lcur,-1.12) -- (\lcur,\bartop+0.95);

\node[lbl, anchor=south] at (\reachlim,\bartop+0.95) {Reach limit};
\node[lbl, anchor=south] at (\tcur,\bartop+0.95)     {Trailing cursor};
\node[lbl, anchor=south] at (\lcur,\bartop+0.95)     {Leading cursor};

\node[lbl, anchor=south] (dead) at (1.42,\bartop+0.40) {Dead thunk groups};
\draw[thin] (dead.south) ++(-0.35,0) -- ({\lcur-7*\dx+0.5*\tw},\bartop+0.06);
\draw[thin] (dead.south) ++( 0.35,0) -- ({\lcur-6*\dx+0.5*\tw},\bartop+0.06);

\node[lbl, anchor=south] (live) at (5.35,\bartop+0.40) {Live thunk groups};
\draw[thin] (live.south) ++(-0.35,0) -- ({\lcur-5*\dx+0.5*\tw},\bartop+0.06);
\draw[thin] (live.south) ++( 0.35,0) -- ({\lcur-4*\dx+0.5*\tw},\bartop+0.06);

\draw[-{Latex[length=2.4mm]}, semithick]
  (8.75,\bartop+0.10) .. controls (10.1,\bartop+1.05) and (12.35,\bartop+1.05)
  .. (\lcur+0.11,\bartop+0.10);
\node[lbl, anchor=south] at (11.3,\bartop+0.80) {Redirected call};

\draw[-{Latex[length=2mm]}, semithick, dash pattern=on 2pt off 1.6pt]
  ({\lcur+\tw-0.03},\bartop+0.14) arc[start angle=170, end angle=10, x radius=0.355, y radius=0.58];

\node[lbl, align=center, anchor=north] (curgrp) at (15.55,-0.28)
  {Thunk group for\\the current batch};
\draw[thin] (curgrp.north) ++(-0.35,0.02) -- ({\lcur+0.5*\tw},\barbot-0.06);

\draw[meas] (\reachlim,-0.45) -- (\batchend,-0.45);
\node[fill=white, inner sep=2pt] at ({(\reachlim+\batchend)/2},-0.45) {Branch reach};

\draw[meas] (\tcur,-0.80) -- (\lcur,-0.80);
\node[fill=white, inner sep=2pt] at ({(\tcur+\lcur)/2},-0.80) {Branch reach minus margin};

\def\reglaby{-1.60}
\node[anchor=base] at (1.4,\reglaby-0.17)                    {Out of reach};
\node[anchor=base] at ({(\reachlim+\tcur)/2},\reglaby-0.17)  {Scanned};
\node[anchor=base] at ({(\tcur+\batchend)/2},\reglaby)      {Current batch,};
\node[anchor=base] at ({(\tcur+\batchend)/2},\reglaby-0.34) {scanning relocations};
\node[anchor=base] at ({(\batchend+\lcur)/2+0.4},\reglaby)      {Address assigned};
\node[anchor=base] at ({(\batchend+\lcur)/2+0.4},\reglaby-0.34) {but not yet scanned};
\node[anchor=base] at (15.7,\reglaby-0.17)                   {No address yet};

\end{tikzpicture}
\caption{Linear-scan thunk creation in \mold. Addresses are assigned up to
the leading cursor, and relocations are scanned in batches at the trailing
cursor, which stays less than one branch reach behind. Each batch emits one
thunk group at the leading cursor. A thunk group stays live
until the scan has advanced one branch reach past it.}
\Description{A horizontal sequence of input sections is divided into
out-of-reach, scanned, current-batch, address-assigned, and unassigned
regions. A trailing cursor scans relocations less than one branch reach
behind a leading cursor that assigns addresses and emits thunk groups.
The diagram also shows live and dead thunk groups and a call redirected
to the current batch's thunk group.}
\label{fig:thunks}
\end{figure*}
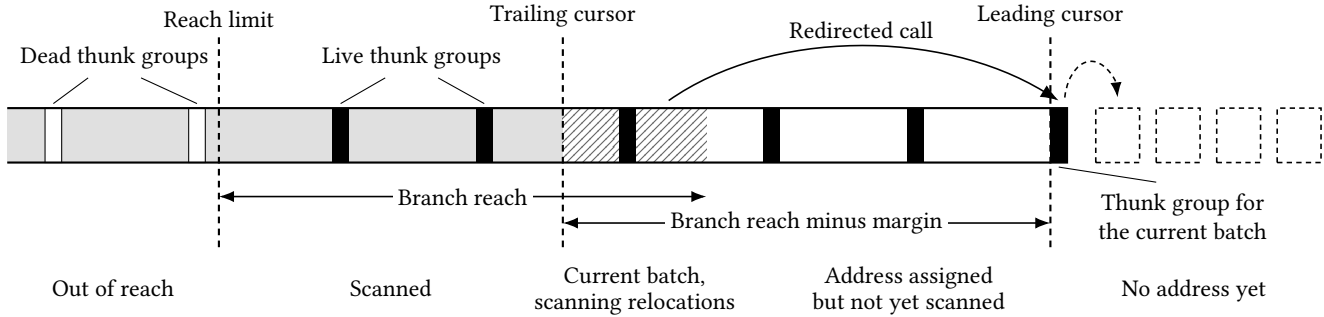

\subsection{File Layout Computation}
\label{sec:layout}

Before copying input section contents to the output file, the linker must assign an output file offset
to every input section. This is essentially a prefix sum over
section sizes, but with two complications: input sections have alignment
requirements, and a single output section may
contain millions of input sections, making a sequential scan a
bottleneck.

\mold uses the standard two-level parallel
scan~\cite{blelloch1990prefix}. First, input sections within
each output section are split into groups of roughly 10,000. The
total size and alignment of each group are computed in parallel.
Second, group offsets within the output section are assigned by a
serial scan over these groups, which are far fewer than the
individual sections. Third, input section
offsets within each group are assigned in another parallel pass, now
that each group's base offset is known. Finally, file offsets for
the output sections themselves are assigned serially, which is trivial
because a typical link produces only a few dozen output sections.

Executable sections on architectures that need range extension
thunks are laid out differently, as described next.

\subsection{Range Extension Thunks}
\label{sec:thunks}

On some architectures, the linker must insert range extension thunks
(\S\ref{sec:concepts}) into the code segment when a call's
displacement exceeds the instruction's immediate field. The challenge is that inserting a thunk
makes the code larger, which may push previously in-range calls
out of range, requiring yet more thunks.

\gnuld, \gold, and \lld all create thunks with essentially the same
sequential fixed-point algorithm: lay out the code first, then
insert thunks into it. \lld, for example, pre-allocates empty thunk
groups at regular intervals and scans all relocations in a single
thread~\cite{smith2017thunks,song2026branches}. When a call turns out to be out of
range, \lld adds a thunk to the lowest-addressed group reachable
from the call. Each insertion, however, lengthens that group and
pushes everything after it to higher addresses. A call that was
scanned earlier and left as a direct branch may have its caller and
callee on opposite sides of that group; its displacement has just
grown, and it may no longer be in range. A single scan therefore
does not suffice: \lld rescans all relocations, inserting more
thunks as needed, until a scan finds nothing new, which usually
takes a few iterations.

\mold eliminates the repetition by creating thunks in a single
linear scan during layout
rather than after it. \revised{Figure~\ref{fig:thunks}\revpoint{C: Fig.~\ref*{fig:thunks}} illustrates the
algorithm.} Within each output section, \mold assigns
addresses to input sections in increasing order, while maintaining
two cursors over the input section list, one leading and one trailing. The
leading cursor points to the end of the section that has just received an
address, the highest assigned so far. The trailing cursor points to the sections whose relocations
are being scanned. The gap between the two cursors is one branch
reach (e.g., 128\,MiB on ARM64) minus a small margin reserved for the
thunks themselves. The scan decides whether each call needs a
thunk. A call that is out of range, or whose target has no address
yet, is redirected to a live thunk for its target if one exists;
otherwise a new thunk is added to the group at the leading cursor
and the call is redirected there.
Inserting a new thunk cannot invalidate any earlier decision, because
every call scanned so far has both its caller and its destination
(the callee or an earlier thunk for it) at lower addresses
than the new thunk.

\mold advances the trailing cursor in batches
so that each batch's relocations can be scanned in parallel.
Entries are deduplicated with a per-symbol atomic flag.

During the scan, calls into other output sections are
pessimistically considered out of range. This assumption is an
optimization: it makes every
output section self-contained, so each one's internal layout can
be computed independently of the others.
This pessimism can overallocate thunk entries. Once the global layout
is known, \mold rescans the relocations in parallel and removes a
symbol from every thunk group if no call to it actually needs a thunk.
Section sizes and offsets are then recomputed.
This grow-then-shrink order is safe: removing entries can only
leave a displacement unchanged or make it smaller, whereas
inserting entries at this stage could make an existing branch
overflow.

\revised{Even though \mold's algorithm is simpler and computationally
cheaper, its
output is comparable to that of the fixed-point
algorithm: for the ARM64 Firefox debug build,
\mold emits\revpoint{B: thunk size} 19,080 thunk entries and \lld
18,896. The small discrepancy, about 2 KiB in a 282 MiB text segment,
is due to minor differences in output layout between the two linkers.}

\subsection{GDB Index Construction}
\label{sec:gdbindex}

A \code{.gdb\_index} section indexes symbol names and address
ranges in DWARF~\cite{dwarfspec} debug information, allowing GDB to
start quickly without scanning all of the DWARF data.
When \code{-gdb-index} is specified, \mold constructs the index.

The index was originally generated by GDB's post-link tool
\code{gdb-add-index}~\cite{gdbindex}, which reads a completed binary and adds the
section. Linkers later gained the ability to
construct it during linking, but the dependency remained one-way.
Index construction consumes what the linker has generated, and no
linker pass consumes the index. \mold therefore builds it
concurrently with other passes, the only major exception to its
serial-pass structure (\S\ref{sec:design}).

\subsection{Output Generation}

In the final phase, \mold memory-maps the output file and processes
all input sections in parallel: each input section copies its data to
the output buffer and applies its relocations in place. This is naturally parallelizable because
each input section writes to a disjoint region of the output.

The last piece of the output is the build ID. ELF executables
commonly contain a \code{.note.gnu.build-id} section that
uniquely identifies the binary; the linker is free to choose any
method to compute the identifier as long as it is unique, and a
cryptographic hash of the output file is typically used. \mold
parallelizes the hash computation using a two-level Merkle
tree~\cite{merkle1988}. It computes a first-level hash over each
4\,MiB block of the output file in parallel, and then hashes the
concatenation of the block hashes to produce the final
identifier.
For the hash function, \mold uses BLAKE3~\cite{blake3}, a
cryptographic hash function designed to take advantage of SIMD
instructions on modern processors.
Hashing the 2.5\,GiB Firefox debug output takes 23\,ms on our
test machine (\S\ref{sec:setup}).

\subsection{Deterministic Output}

A parallel linker can easily produce nondeterministic output.
For example, when multiple threads insert strings into a shared string
pool concurrently, the order of strings within the pool varies
with thread scheduling; a linker that assigns output offsets to the strings
in pool order will then emit them in a different order on each run.
Each ordering is correct---the linked binary works regardless---but
the output is no longer bit-identical across runs, breaking build
reproducibility~\cite{reproduciblebuilds}.

\mold avoids this by designing each parallel phase so that its result
does not depend on the order in which threads execute. This is
achieved either by using commutative operations (e.g., setting atomic
flags on symbols) or by following a nondeterministic parallel step
with a pass that sorts its results into a canonical order. As a result, \mold guarantees that
the same inputs, command-line options, and linker version always
produce a bit-identical binary.

\section{System-Level Optimizations}
\label{sec:io}

Beyond the parallel pipeline described above, \mold employs several
system-level optimizations to reduce link time.

\subsection{Huge Pages for the Output File}
\label{sec:hugepage}

\mold writes the output file through a memory mapping. On Linux,
every page of a fresh output file is created on first touch by
a minor page fault, which adds up to more than a million faults for
a multi-gigabyte output, and concurrent fault handling contends on
per-file kernel locks. \mold therefore asks the kernel to back the
output mapping with huge pages using
\code{madvise(MADV\_HUGEPAGE)}. On kernels that can use huge pages
for file-backed memory, as recent Linux kernels can for common
filesystems such as ext4, the output is then faulted in huge-page
units rather than base pages (e.g., 2\,MiB versus 4\,KiB on x86-64),
cutting the fault count by orders of magnitude. This single
\code{madvise} call makes \mold's output copy pass
3.1x faster and shortens the Firefox debug link from 1.46\,s to
0.89\,s.

The optimization is easy for any linker to adopt, which raises the
question of how much of \mold's advantage it explains. The answer is
governed by Amdahl's law. We applied the same one-line change to
\lld, and its time for the same link improved by 12\% (7.04\,s
to 6.19\,s), versus 39\% for \mold. Output copy accounts for about
a quarter of \lld's largely sequential runtime, so even a large
improvement to that pass moves the total only so far. A system-level
optimization like this pays off in proportion to how little other
work remains, and pervasive parallelism leaves \mold with far less
of it than \lld.

\subsection{Output File Block Preallocation}
\label{sec:fallocate}

A freshly created output file is \emph{sparse}. That is,
\code{ftruncate} sets its size, but no disk space is assigned
until the file is actually written. The first write to each page
of the mapped output therefore enters the page-fault handler,
which must reserve disk space for the page. The
filesystem in our evaluation, ext4, serializes these reservations on
a per-file lock, so dozens of threads writing a multi-gigabyte
output spend more time spinning on the lock than copying.
\mold avoids this by preallocating the
entire file with one \code{fallocate} system call before mapping
it; preallocating extents for the 2.5\,GiB Firefox debug output takes
about 5\,ms. The call makes \mold's copy pass 3.5x faster (0.82\,s
to 0.24\,s) and shortens the Firefox debug link from 1.46\,s to
0.89\,s, coincidentally the same figures as in
\S\ref{sec:hugepage}. Consistent with the proportionality rule of
\S\ref{sec:hugepage}, the same one-line change applied to \lld
shortens the same link by only 3.8\% (7.04\,s to 6.77\,s).

\subsection{Memory Allocator}
\label{sec:allocator}

A parallel linker stresses the memory allocator, as dozens of
threads allocate and free small objects concurrently. \mold uses
mimalloc~\cite{leijen2019mimalloc}, an allocator designed for
concurrent workloads. The choice is worth a factor of
1.3: linked against the glibc malloc instead, \mold
takes 1.15\,s rather than 0.89\,s on the Firefox debug build.
Other scalable allocators land in between:
jemalloc~\cite{evans2006jemalloc} at 1.03\,s,
tcmalloc~\cite{ghemawat2007tcmalloc} at 1.09\,s, and
tbbmalloc~\cite{onetbb} at 1.12\,s.

To gauge how close mimalloc comes to the practical limit for this
workload, we also implemented a simple bump-pointer allocator that
allocates from large per-thread buffers and never frees individual
objects. It was in fact slower (1.00\,s) and increased peak memory
consumption by 2\%. This result suggests that replacing
mimalloc with another general-purpose allocator is unlikely to yield a
substantial further speedup.

Switching from the glibc malloc to mimalloc helps \lld too,
shortening its time on the same workload from 7.04\,s to 5.97\,s.
Memory allocation, unlike output
copy, occurs in every pass, so by the proportionality rule of
\S\ref{sec:hugepage}, \lld gains more from the allocator than from
either output file optimization.

\subsection{Output File Reuse}
\label{sec:reuse}

In an edit-compile-debug cycle, the output file the
linker is about to write usually already exists from the previous
build. Writing a new file and renaming it over the old one
implicitly deletes the old file, and deleting a multi-gigabyte
file is not cheap: the kernel must tear down its page cache and
release its disk blocks.
\mold therefore overwrites the existing file in place when it can,
reusing its inode, its disk blocks, and its page cache. Linux
refuses to open a running executable for writing, so overwriting
cannot corrupt a program that is currently running; if the open
fails, \mold falls back to creating a new file. Shared libraries
have no such protection, and a running process may have the old
file mapped, so for shared library outputs \mold always creates a
new file. Firefox's main binary is a shared
library, so we measure this optimization on Chromium's 4.5\,GiB
debug executable instead: overwriting saves about 0.3\,s.

\subsection{Two-Process Architecture}
\label{sec:fork}

When a linker calls \code{\_exit}, the kernel reclaims all the
resources the process has been holding, tearing down its address
space in a single thread, and notifies the parent of the
termination only after that. Since a linker is a large process
with many memory-mapped files, the
reclamation takes an appreciable amount of time. The parent is the
build system waiting for the linker to finish, so that time is
perceived as link time. \mold
reduces the teardown cost by dropping its file mappings in parallel
with \code{madvise(MADV\_DONTNEED)} before exiting (unlike
\code{munmap}, \code{madvise} does not serialize on the
exclusive address-space lock on Linux); the contents remain in the page
cache, so the next link still finds the files warm. The technique
is linker-independent: retrofitted into \lld (\S\ref{sec:setup}), it
shortens the Firefox debug link by 0.3\,s.

The remaining teardown latency is hidden by a two-process
architecture. At startup, \mold forks a child process that
performs the actual linking while the original process waits.
When the child finishes writing and closing the output file, it
signals the waiting process, which exits immediately, returning
control to the build system; the parent's teardown is negligible
because it holds almost no resources.
The child then exits, and the kernel reclaims its resources in the
background, invisible to the user. On top of the exit cost
reduction described above, the two-process architecture saves a
further 10\% of the user-perceived link time,
for example 0.10\,s for a Firefox debug link.

\section{Implementation}
\label{sec:impldetails}

\mold is implemented in approximately 28,000 lines of C++20 code
shared across all targets, plus 300--1300 lines per target
architecture. It uses oneTBB for parallel execution.

\subsection{Compact Metadata Representation}
\label{sec:memory}

\mold keeps its in-memory linker metadata compact to reduce memory traffic
and cache misses. Objects such as input files, symbols, and
sections are allocated in a dedicated arena and refer to one another
using 32-bit offsets within the arena instead of 64-bit pointers.
This halves the storage required for each cross-reference and reduces
the size of metadata records traversed by many linker passes.

\subsection{Architecture-Parametric Design}

\mold supports 14 processor architectures (i386, x86-64, ARM 32/64, RISC-V 32/64, PowerPC 32/64,
LoongArch 32/64, s390x, SPARC64, m68k, and SH-4) through C++ templates. The core linker logic is
parameterized on an architecture type \code{E} that provides types
for ELF data structures (e.g., headers, symbols, relocations) and
constants (e.g., relocation type enumerations, page sizes).
Architecture-specific code is implemented in separate files
(\code{arch-*.cc}).

This design avoids the overhead of virtual dispatch
at the cost of a larger linker binary, since the core logic is
instantiated once per target architecture.

\subsection{Link-Time Optimization Support}
\label{sec:lto}

\mold supports GCC and LLVM Link-Time Optimization (LTO). At an
early stage of the linking pipeline, \mold loads the respective
compiler's LTO plugin via \code{dlopen} and delegates
intermediate-representation compilation to it, producing native object
files that are then linked normally. When LTO is enabled, the
plugin's compilation step dominates the end-to-end link time, so \mold's speed
advantage is diminished. In practice this matters little, because LTO is typically reserved
for optimized release builds, not the debug builds where linking
speed is most critical.

\section{Performance Evaluation}
\label{sec:eval}

We evaluate \mold's performance along four dimensions: end-to-end
performance on real-world workloads (\S\ref{sec:eval-e2e}),
scalability and CPU cost (\S\ref{sec:eval-scale}), an ablation study
(\S\ref{sec:eval-ablation}), and a per-pass comparison with \lld
(\S\ref{sec:eval-lld}--\S\ref{sec:eval-thunks}).

\subsection{Experimental Setup}
\label{sec:setup}

Experiments were conducted on two machines. The primary machine
has an AMD Ryzen Threadripper 7980X (64 cores, 128 threads),
384\,GiB of DDR5 RAM, running Linux 6.17 on Ubuntu 24.04.
The second machine is an Apple Mac Studio with an M1 Ultra SoC
(16 performance cores and 4 efficiency cores) and 128\,GiB of
unified LPDDR5 memory, running Fedora Asahi Remix 42 with
Linux 6.18. On both machines, the benchmarks reside on an ext4
filesystem on NVMe storage. Frequency boost is disabled, the
frequency governor is pinned to performance, and transparent huge
pages are in madvise mode. Unless
otherwise noted, results are from the x86-64 machine. We compare four linkers:

\begin{itemize}
\item \mold \moldversion
\item \lld 22.1.8, with \mold's system-level optimizations
  retrofitted (see below)
\item \gold 2.46.1
\item \gnuld 2.46.1
\end{itemize}

\noindent
The system-level optimizations of \S\ref{sec:io} require care in how they
enter the comparison. Adopting huge pages (\S\ref{sec:hugepage}),
file preallocation (\S\ref{sec:fallocate}), the memory allocator
(\S\ref{sec:allocator}), and the exit-time teardown of file
mappings (\S\ref{sec:fork}) requires no architectural
restructuring, so they are
available to other linkers as-is. To separate
their effect from \mold's architecture, we modify \lld to use all
four and measure it in this retrofitted configuration
throughout; the reported gaps then reflect architectural
differences.
The four optimizations compose roughly additively, making \lld
1.1--2.1x faster than as shipped across our benchmarks; the
Firefox debug link, for example, shortens
from 7.04\,s to 4.44\,s.
\gnuld and \gold are measured
unmodified: they are one to two orders of magnitude behind on every
workload, and the retrofit would not change that picture.

\revised{The other two optimizations of \S\ref{sec:io} depend on conditions
a benchmark must control: output file reuse helps only when a
previous output exists on disk, and the two-process architecture
does not speed up linking but hides the process teardown latency
from the build system. We disable both, by removing the
output file before each run and by passing \code{-no-fork}\revpoint{B-Q2} to
\code{\mold}, so
that every linker is measured on the same task: building the
output from scratch and terminating.}

We selected workloads from nine large, widely used open-source
programs.
Table~\ref{tab:workloads} lists them along with their input and
output sizes. \revised{We measure the time to link each program's main
binary in two configurations:\revpoint{A-W2} a \emph{debug} build with debug
information and a \emph{release} build without it.}
These produce output binaries ranging from 0.15\,GiB to
9.91\,GiB, making link time a meaningful fraction of the
overall build cycle.

\begin{table}
\caption{Benchmark workloads and their input and output sizes.
  Sizes are in GiB.}
\label{tab:workloads}
\centering
\small

\begingroup
\sisetup{
  group-separator = {,},
  group-minimum-digits = 4,
  table-number-alignment = right,
}
\setlength{\tabcolsep}{6pt}

\begin{tabular}{
  @{}l
  S[table-format=5.0]
  S[table-format=2.2]
  S[table-format=1.2]
  S[table-format=2.2]
  S[table-format=1.2]
  @{}
}
\toprule
\multirow[b]{2}{*}[
  -\dimexpr\aboverulesep+\cmidrulewidth+\belowrulesep\relax
]{Program}
&
\multicolumn{1}{c}{%
  \multirow[b]{2}{*}[
    -\dimexpr\aboverulesep+\cmidrulewidth+\belowrulesep\relax
  ]{\makecell[b]{Number of\\object files}}%
}
& \multicolumn{2}{c}{Release}
& \multicolumn{2}{c}{Debug} \\
\cmidrule(lr){3-4}
\cmidrule(l){5-6}
& & {In} & {Out} & {In} & {Out} \\
\midrule
Blender 5.2      &  9943 & 0.90 & 0.25 &  8.75 & 2.52 \\
Chromium 145     & 40945 & 2.20 & 0.53 & 12.31 & 4.53 \\
Clang 21.1       &  3030 & 0.45 & 0.21 & 18.11 & 4.25 \\
ClickHouse 26.1  & 14280 & 3.17 & 1.21 & 12.82 & 5.59 \\
Firefox 149      &  3155 & 0.60 & 0.22 &  6.48 & 2.47 \\
Godot 4.6        &  2682 & 0.46 & 0.15 &  4.40 & 1.11 \\
LibreOffice 26.2 &  7185 & 0.58 & 0.19 &  4.07 & 1.00 \\
PyTorch 2.9      &  2793 & 0.69 & 0.32 &  9.87 & 3.60 \\
TensorFlow 2.21  &  8420 & 1.83 & 0.81 & 23.87 & 9.91 \\
\bottomrule
\end{tabular}

\endgroup
\end{table}

Link time matters most for large debug builds, which developers relink
repeatedly in the edit-compile-debug cycle. Speeding up such
links is \mold's primary design goal.
The detailed analyses in \S\ref{sec:eval-scale}--\S\ref{sec:eval-lld}
use the Firefox debug build as a running example.
We do not evaluate link-time optimization workloads because LTO is
typically used only for production builds, where the LTO compilation
step itself dominates total time and \mold's advantage is negligible
(\S\ref{sec:lto}).

Each measurement is the median of five runs after one warmup run
that brings input files into the page cache, reflecting the
realistic scenario in which object files remain cached from the
immediately preceding compilation.

\subsection{End-to-End Performance}
\label{sec:eval-e2e}

Table~\ref{tab:e2e} shows end-to-end link times for the release and
debug configurations of each program, with every linker at its
default thread count.
Speedups generally grow with link size. Small links are dominated
by startup costs that both linkers pay alike, and as links grow,
more of \lld's time goes to its single-threaded passes
(\S\ref{sec:eval-lld}). The largest speedup, 16.1x on the
TensorFlow debug build, combines size with a workload-specific
cause: the link applies a version script with two dozen glob
patterns, which \lld matches against 2.5 million defined symbols
one pattern at a time on a single thread. That pass takes about
30 of \lld's 52 seconds. Even without it, the speedup would be
about 7x. The second largest, Chromium's debug build at 7.0x, is
a pure case of size: its link has by far the most input files in
the suite (40,945; Table~\ref{tab:workloads}), and \lld spends
about 7 of its 13.2 seconds parsing them. The other programs gain
2.4--5.7x.

\begin{table}[t]
\centering
\caption{End-to-end link times in seconds; for each program, the
  first row is the release build and the second row the debug build.
  Lower is better. Speedup is \mold's speedup over \lld. A ``---''
  indicates that the linker cannot link the workload. Here and in
  all following tables and figures, \lld denotes the retrofitted
  configuration described in \S\ref{sec:setup}.}
\label{tab:e2e}
\small
\begin{tabular}{@{}lrrrrr@{}}
\toprule
Program & \mold & \lld & \gold & \gnuld & Speedup \\
\midrule
Blender     & 0.22 &  0.82 &  5.14 & 11.32 & 3.7x \\
            & 0.98 &  4.14 & 55.17 & 74.40 & 4.2x \\
\addlinespace
Chromium    & 1.05 &  6.29 & ---   & ---   & 6.0x \\
            & 1.89 & 13.24 & ---   & ---   & 7.0x \\
\addlinespace
Clang       & 0.13 &  0.37 &  3.15 &  6.66 & 2.8x \\
            & 1.48 &  4.54 & 91.31 & 81.84 & 3.1x \\
\addlinespace
ClickHouse  & 0.47 &  2.66 & ---   & ---   & 5.7x \\
            & 1.20 &  4.87 & ---   & ---   & 4.1x \\
\addlinespace
Firefox     & 0.25 &  0.98 & ---   & 14.33 & 3.9x \\
            & 0.89 &  4.44 & ---   & 92.86 & 5.0x \\
\addlinespace
Godot       & 0.11 &  0.34 &  2.49 &  6.02 & 3.1x \\
            & 0.51 &  1.20 & 20.56 & 26.64 & 2.4x \\
\addlinespace
LibreOffice & 0.26 &  0.98 &  3.09 & 29.13 & 3.8x \\
            & 0.46 &  2.50 & 23.20 & ---   & 5.4x \\
\addlinespace
PyTorch     & 0.17 &  0.47 &  3.49 &  7.47 & 2.8x \\
            & 0.88 &  2.60 & 47.65 & 59.97 & 3.0x \\
\addlinespace
TensorFlow  & 0.75 & 10.45 & ---   & ---   & 13.9x \\
            & 3.23 & 52.16 & ---   & ---   & 16.1x \\
\bottomrule
\end{tabular}
\end{table}

The empty cells in Table~\ref{tab:e2e} mark configurations that a linker cannot
link. The GNU linkers fail on different workloads for different
reasons, each of which falls into one of four categories: the workload
uses a command-line option implemented only by \lld and \mold, relies on
their order-insensitive archive symbol resolution (\S\ref{sec:symres}), references
a symbol synthesized only by them, or uses a new ELF feature that the
GNU linkers do not yet support.
That 7 of our 18 workload configurations cannot be linked by \gnuld
suggests that, for large modern C++ projects, \lld rather
than \gnuld has become the practical compatibility baseline.

\begin{table}[t]
\centering
\caption{Peak memory usage (maximum resident set size) in GiB
  when linking the debug builds.}
\label{tab:rss}
\small
\begin{tabular}{@{}l
  S[table-format=2.2] S[table-format=2.2]
  S[table-format=2.2] S[table-format=2.2]@{}}
\toprule
Program & {\mold} & {\lld} & {\gold} & {\gnuld} \\
\midrule
Blender     & 15.51 & 15.62 & 18.55 &  9.37 \\
Chromium    & 23.57 & 26.44 & {---} & {---} \\
Clang       & 25.62 & 26.75 & 32.63 & 13.61 \\
ClickHouse  & 21.09 & 22.35 & {---} & {---} \\
Firefox     & 11.24 & 12.07 & {---} & 10.64 \\
Godot       &  6.69 &  6.89 &  8.48 &  4.37 \\
LibreOffice &  6.57 &  7.40 &  6.82 & {---} \\
PyTorch     & 15.41 & 16.43 & 19.38 &  9.86 \\
TensorFlow  & 37.35 & 40.60 & {---} & {---} \\
\bottomrule
\end{tabular}
\end{table}

\revised{Table~\ref{tab:rss}\revpoint{C-Q3} lists peak memory
usage on the debug builds. \mold, \lld, and \gold perform I/O
through \code{mmap} and reach similar peaks because the
dominant term is one they share: the resident pages of the
mapped input and output files. \mold's eager parsing of every archive
member adds little, because it reads only the member's symbol and
section tables, not the section contents.
\gnuld performs I/O with explicit \code{read}
and \code{write} calls, and its peak RSS on the large debug
links is therefore markedly lower.}

\subsubsection{ARM64 Results}

Table~\ref{tab:e2e-arm64} shows end-to-end link times on the
Apple M1 Ultra, an ARM64 machine with 16 performance cores
(restricted to those cores via \code{taskset}).
The speedup of \mold over \lld is 1.7--12.6x, lower than the
2.4--16.1x on x86-64. Relative to the Threadripper, the M1 offers
fewer but faster cores, a trade that favors a sequential linker.
\lld inherits much of the per-core speedup, linking the
Firefox debug build in 3.26\,s versus 4.43\,s.
\mold, in contrast, loses more to the missing
cores than it gains from the faster ones: the same link takes
1.02\,s with 16 threads on the M1, up from 0.89\,s with 32
threads on the Threadripper.
Nevertheless, the results confirm that \mold's advantage holds
across both machines.

\begin{table}[t]
\centering
\caption{End-to-end link times on ARM64 with 16 cores (seconds).
  Conventions are as in Table~\ref{tab:e2e}.}
\label{tab:e2e-arm64}
\small
\begin{tabular}{@{}lrrrrr@{}}
\toprule
Program & \mold & \lld & \gold & \gnuld & Speedup \\
\midrule
Blender           & 0.24 &  0.59 &  3.37 &  6.17 & 2.5x \\
                  & 1.41 &  3.39 & 38.47 & 46.29 & 2.4x \\
\addlinespace
Chromium          & 1.04 &  4.70 & ---   & ---   & 4.5x \\
                  & 2.43 &  9.97 & ---   & ---   & 4.1x \\
\addlinespace
Clang             & 0.15 &  0.26 &  1.92 &  3.22 & 1.7x \\
                  & 2.37 &  3.97 & 64.91 & 53.61 & 1.7x \\
\addlinespace
ClickHouse        & 0.61 &  1.94 & ---   & ---   & 3.2x \\
                  & 1.71 &  4.01 & ---   & ---   & 2.3x \\
\addlinespace
Firefox           & 0.21 &  0.60 & ---   &  7.04 & 2.9x \\
                  & 1.02 &  3.26 & ---   & 50.49 & 3.2x \\
\addlinespace
Godot             & 0.11 &  0.24 &  1.55 &  3.04 & 2.2x \\
                  & 0.59 &  1.00 & 14.56 & 15.88 & 1.7x \\
\addlinespace
LibreOffice       & 0.25 &  0.55 &  2.01 & 13.83 & 2.2x \\
                  & 0.66 &  1.86 & 17.69 & ---   & 2.8x \\
\addlinespace
PyTorch           & 0.18 &  0.33 &  2.30 &  4.12 & 1.8x \\
                  & 1.31 &  2.27 & 34.92 & 38.15 & 1.7x \\
\addlinespace
TensorFlow        & 0.71 &  8.96 & ---   & ---   & 12.6x \\
                  & 4.32 & 47.09 & ---   & ---   & 10.9x \\
\bottomrule
\end{tabular}
\end{table}

\subsection{Scalability}
\label{sec:eval-scale}

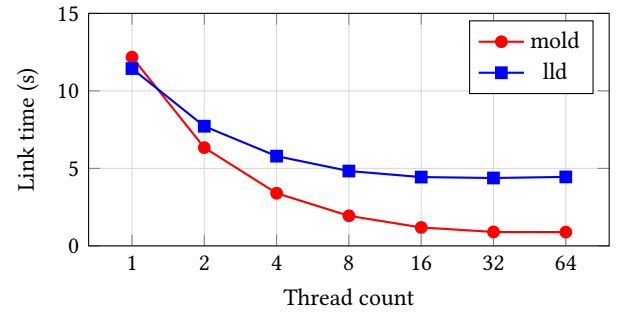
\begin{figure}[t]
\centering
\begin{tikzpicture}
\begin{axis}[
  width=\columnwidth,
  height=0.55\columnwidth,
  xlabel={Thread count},
  ylabel={Link time (s)},
  xmode=log,
  log basis x=2,
  xtick={1,2,4,8,16,32,64},
  xticklabels={1,2,4,8,16,32,64},
  ymin=0,
  ymax=15,
  legend pos=north east,
  legend style={font=\small},
  grid=major,
  grid style={gray!30},
  tick label style={font=\small},
  label style={font=\small},
]
\addplot[thick, mark=*, red] coordinates {
  (1,12.18) (2,6.34) (4,3.40) (8,1.94) (16,1.19) (32,0.90) (64,0.89)
};
\addlegendentry{\mold}
\addplot[thick, mark=square*, blue] coordinates {
  (1,11.44) (2,7.72) (4,5.79) (8,4.83) (16,4.44) (32,4.38) (64,4.45)
};
\addlegendentry{\lld}
\end{axis}
\end{tikzpicture}
\caption{Scalability of \mold and \lld on the Firefox debug build.}
\Description{A line chart of link time versus thread count. \mold's time
falls from 12.18 seconds with one thread to about 0.9 seconds with 32
threads and then plateaus. lld's time falls from 11.44 seconds with one
thread to about 4.4 seconds with 16 threads and then plateaus.}
\label{fig:scalability}
\end{figure}

Figure~\ref{fig:scalability} plots the link times of \mold and \lld
for the Firefox debug build as a function of the worker thread
count, from 1 to 64. \mold speeds up steadily with thread count:
1.9x at 2 threads, 6.3x at 8, and 13.5x at 32, beyond which it
plateaus. \lld plateaus earlier, at 16 threads, with only 2.6x
speedup. These results led us to set \mold's default cap at
32 threads and are consistent with \lld's default cap at 16 threads.

At one thread, the two linkers are not far apart (12.2\,s
for \mold vs.\ 11.4\,s for \lld). The near-parity comes from two
opposing effects that roughly cancel. \mold's parallel architecture has costs
that a sequential linker does not pay, such as the eager full
parsing of every archive member (\S\ref{sec:impl-parse}) and the
extra rounds of work some passes trade for parallelism
(\S\ref{sec:layout}). Offsetting this, \mold employs faster algorithms
in some passes and utilizes compact data representations
(\S\ref{sec:memory}). The balance tips
decisively as threads are added, confirming that \mold's
advantage comes from parallel scalability.

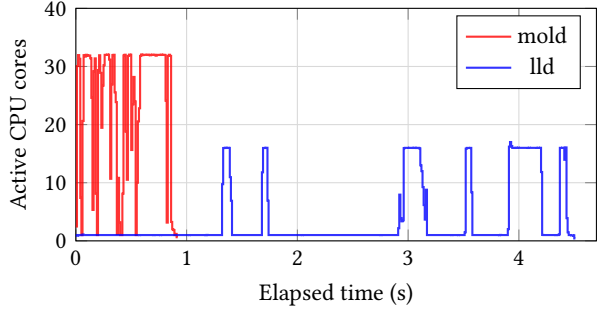
\begin{figure}[t]
\centering
\begin{tikzpicture}
\begin{axis}[
  width=\columnwidth,
  height=0.55\columnwidth,
  xlabel={Elapsed time (s)},
  ylabel={Active CPU cores},
  xmin=0, xmax=4.7,
  ymin=0, ymax=40,
  legend pos=north east,
  legend style={font=\small},
  grid=major,
  grid style={gray!30},
  tick label style={font=\small},
  label style={font=\small},
]
\addplot[thick, red, opacity=0.8, const plot] coordinates {
  (0.00,2.47) (0.01,30.07) (0.02,31.96) (0.03,31.32) (0.04,7.38) (0.05,1.00)
  (0.06,20.63) (0.07,31.95) (0.08,31.70) (0.09,32.01) (0.10,31.94) (0.11,31.83)
  (0.12,31.89) (0.13,31.42) (0.14,24.60) (0.15,5.32) (0.16,29.19) (0.17,31.39)
  (0.18,8.12) (0.19,1.00) (0.20,29.38) (0.21,31.93) (0.22,28.90) (0.23,19.45)
  (0.24,26.71) (0.25,30.21) (0.26,32.00) (0.27,32.00) (0.28,32.05) (0.29,32.03)
  (0.30,31.18) (0.31,10.68) (0.32,26.91) (0.33,31.97) (0.34,32.01) (0.35,25.55)
  (0.36,23.87) (0.37,1.18) (0.38,10.84) (0.39,3.19) (0.40,1.00) (0.41,1.00)
  (0.42,8.12) (0.43,31.89) (0.44,32.02) (0.45,18.80) (0.46,10.51) (0.47,31.82)
  (0.48,32.04) (0.49,32.03) (0.50,15.63) (0.51,23.49) (0.52,28.20) (0.53,23.99)
  (0.54,1.00) (0.55,10.00) (0.56,20.55) (0.57,22.81) (0.58,31.82) (0.59,32.07)
  (0.60,31.95) (0.61,32.00) (0.62,31.99) (0.63,31.99) (0.64,32.04) (0.65,32.04)
  (0.66,31.96) (0.67,31.93) (0.68,32.02) (0.69,32.03) (0.70,32.00) (0.71,31.98)
  (0.72,31.86) (0.73,32.00) (0.74,32.00) (0.75,31.98) (0.76,31.98) (0.77,32.04)
  (0.78,31.99) (0.79,32.00) (0.80,31.98) (0.81,24.11) (0.82,3.14) (0.83,31.82)
  (0.84,32.01) (0.85,31.82) (0.86,3.25) (0.87,2.00) (0.88,1.86) (0.89,1.00)
  (0.90,1.20) (0.91,0.63)
  (0.92,0.63)
};
\addlegendentry{\mold}
\addplot[thick, blue, opacity=0.8, const plot] coordinates {
  (0.00,1.00) (0.01,0.89) (0.02,1.00) (0.03,1.00) (0.04,0.99) (0.05,1.00)
  (0.06,1.09) (0.07,1.00) (0.08,1.00) (0.09,1.00) (0.10,1.00) (0.11,1.00)
  (0.12,1.00) (0.13,1.00) (0.14,1.00) (0.15,1.00) (0.16,1.00) (0.17,1.00)
  (0.18,1.00) (0.19,1.00) (0.20,1.00) (0.21,1.00) (0.22,1.00) (0.23,1.00)
  (0.24,1.00) (0.25,1.00) (0.26,1.00) (0.27,1.00) (0.28,1.00) (0.29,1.00)
  (0.30,1.00) (0.31,1.00) (0.32,1.00) (0.33,1.00) (0.34,1.00) (0.35,1.00)
  (0.36,1.00) (0.37,1.00) (0.38,1.00) (0.39,1.00) (0.40,1.00) (0.41,1.00)
  (0.42,1.00) (0.43,1.00) (0.44,1.00) (0.45,1.00) (0.46,1.00) (0.47,1.00)
  (0.48,1.00) (0.49,1.00) (0.50,1.00) (0.51,1.00) (0.52,1.00) (0.53,1.00)
  (0.54,1.00) (0.55,1.00) (0.56,1.00) (0.57,0.99) (0.58,1.00) (0.59,1.00)
  (0.60,1.00) (0.61,1.00) (0.62,1.00) (0.63,1.00) (0.64,1.00) (0.65,1.00)
  (0.66,1.00) (0.67,1.00) (0.68,1.00) (0.69,1.00) (0.70,0.99) (0.71,1.00)
  (0.72,1.00) (0.73,1.00) (0.74,1.00) (0.75,1.00) (0.76,0.99) (0.77,1.00)
  (0.78,1.00) (0.79,1.00) (0.80,1.00) (0.81,1.00) (0.82,1.00) (0.83,1.00)
  (0.84,1.00) (0.85,1.00) (0.86,1.00) (0.87,1.00) (0.88,1.00) (0.89,1.00)
  (0.90,1.00) (0.91,1.00) (0.92,1.00) (0.93,1.00) (0.94,1.00) (0.95,1.00)
  (0.96,1.00) (0.97,1.00) (0.98,1.00) (0.99,1.00) (1.00,1.00) (1.01,1.00)
  (1.02,0.99) (1.03,1.00) (1.04,1.00) (1.05,1.00) (1.06,1.00) (1.07,1.00)
  (1.08,1.00) (1.09,1.00) (1.10,1.00) (1.11,1.00) (1.12,1.00) (1.13,1.00)
  (1.14,1.00) (1.15,1.00) (1.16,1.00) (1.17,0.98) (1.18,1.02) (1.19,1.00)
  (1.20,1.00) (1.21,1.00) (1.22,1.00) (1.23,1.00) (1.24,1.00) (1.25,1.00)
  (1.26,1.00) (1.27,1.00) (1.28,1.00) (1.29,1.00) (1.30,1.00) (1.31,1.00)
  (1.32,9.75) (1.33,15.97) (1.34,15.98) (1.35,16.02) (1.36,15.98) (1.37,16.00)
  (1.38,15.99) (1.39,10.96) (1.40,6.90) (1.41,1.00) (1.42,1.00) (1.43,1.00)
  (1.44,1.00) (1.45,1.00) (1.46,1.00) (1.47,1.00) (1.48,1.00) (1.49,1.00)
  (1.50,1.00) (1.51,1.00) (1.52,1.00) (1.53,1.00) (1.54,1.00) (1.55,1.00)
  (1.56,1.00) (1.57,1.00) (1.58,1.00) (1.59,1.00) (1.60,1.00) (1.61,1.00)
  (1.62,1.00) (1.63,1.00) (1.64,1.00) (1.65,1.00) (1.66,1.00) (1.67,1.00)
  (1.68,11.79) (1.69,15.98) (1.70,15.97) (1.71,16.00) (1.72,16.01) (1.73,6.91)
  (1.74,1.00) (1.75,1.00) (1.76,1.00) (1.77,1.00) (1.78,1.00) (1.79,1.00)
  (1.80,1.00) (1.81,1.00) (1.82,1.00) (1.83,1.00) (1.84,1.00) (1.85,1.00)
  (1.86,1.00) (1.87,1.00) (1.88,1.00) (1.89,1.00) (1.90,1.00) (1.91,1.00)
  (1.92,1.00) (1.93,1.00) (1.94,1.00) (1.95,1.00) (1.96,1.00) (1.97,1.00)
  (1.98,1.00) (1.99,1.00) (2.00,1.00) (2.01,1.00) (2.02,1.00) (2.03,1.00)
  (2.04,0.99) (2.05,1.00) (2.06,1.00) (2.07,1.00) (2.08,1.00) (2.09,1.00)
  (2.10,1.00) (2.11,1.00) (2.12,1.00) (2.13,1.00) (2.14,1.00) (2.15,1.00)
  (2.16,1.00) (2.17,1.00) (2.18,1.00) (2.19,1.00) (2.20,1.00) (2.21,1.00)
  (2.22,1.00) (2.23,1.00) (2.24,1.00) (2.25,1.00) (2.26,1.00) (2.27,1.00)
  (2.28,1.00) (2.29,1.00) (2.30,1.00) (2.31,1.00) (2.32,1.00) (2.33,1.00)
  (2.34,1.00) (2.35,1.00) (2.36,1.00) (2.37,1.00) (2.38,1.00) (2.39,1.00)
  (2.40,1.00) (2.41,1.00) (2.42,1.00) (2.43,1.00) (2.44,1.00) (2.45,1.00)
  (2.46,1.00) (2.47,1.00) (2.48,1.00) (2.49,1.00) (2.50,1.00) (2.51,1.00)
  (2.52,1.00) (2.53,1.00) (2.54,1.00) (2.55,1.00) (2.56,1.00) (2.57,1.00)
  (2.58,1.00) (2.59,1.00) (2.60,1.00) (2.61,1.00) (2.62,1.00) (2.63,1.00)
  (2.64,1.00) (2.65,1.00) (2.66,1.00) (2.67,1.00) (2.68,1.00) (2.69,1.00)
  (2.70,1.00) (2.71,1.00) (2.72,1.00) (2.73,1.00) (2.74,1.00) (2.75,0.99)
  (2.76,1.00) (2.77,1.00) (2.78,1.00) (2.79,1.00) (2.80,1.00) (2.81,1.00)
  (2.82,1.00) (2.83,1.00) (2.84,1.00) (2.85,1.00) (2.86,1.00) (2.87,1.00)
  (2.88,1.00) (2.89,1.00) (2.90,1.00) (2.91,2.18) (2.92,7.95) (2.93,3.90)
  (2.94,3.36) (2.95,3.53) (2.96,15.95) (2.97,16.00) (2.98,16.00) (2.99,16.00)
  (3.00,15.99) (3.01,16.00) (3.02,16.00) (3.03,16.00) (3.04,15.98) (3.05,16.01)
  (3.06,16.00) (3.07,15.97) (3.08,16.00) (3.09,16.00) (3.10,15.90) (3.11,13.00)
  (3.12,12.00) (3.13,9.25) (3.14,6.16) (3.15,3.66) (3.16,8.80) (3.17,1.00)
  (3.18,1.00) (3.19,1.00) (3.20,1.00) (3.21,1.00) (3.22,1.00) (3.23,1.00)
  (3.24,1.00) (3.25,1.00) (3.26,1.00) (3.27,1.00) (3.28,1.00) (3.29,1.00)
  (3.30,1.00) (3.31,1.00) (3.32,1.00) (3.33,1.00) (3.34,1.00) (3.35,1.00)
  (3.36,1.00) (3.37,1.00) (3.38,1.00) (3.39,0.99) (3.40,1.00) (3.41,1.00)
  (3.42,1.00) (3.43,1.00) (3.44,1.00) (3.45,0.99) (3.46,1.00) (3.47,1.00)
  (3.48,1.00) (3.49,1.00) (3.50,1.00) (3.51,1.47) (3.52,15.86) (3.53,15.99)
  (3.54,15.99) (3.55,15.97) (3.56,15.99) (3.57,8.36) (3.58,1.00) (3.59,1.00)
  (3.60,1.00) (3.61,1.00) (3.62,1.00) (3.63,1.00) (3.64,1.00) (3.65,1.00)
  (3.66,1.00) (3.67,1.00) (3.68,1.00) (3.69,1.00) (3.70,1.00) (3.71,1.00)
  (3.72,1.00) (3.73,1.00) (3.74,1.00) (3.75,1.00) (3.76,1.00) (3.77,1.00)
  (3.78,1.00) (3.79,1.00) (3.80,1.00) (3.81,1.00) (3.82,0.98) (3.83,1.02)
  (3.84,1.00) (3.85,1.00) (3.86,1.00) (3.87,1.00) (3.88,1.00) (3.89,1.00)
  (3.90,2.02) (3.91,16.21) (3.92,17.00) (3.93,16.31) (3.94,15.99) (3.95,15.99)
  (3.96,15.98) (3.97,16.01) (3.98,16.00) (3.99,16.02) (4.00,16.00) (4.01,16.00)
  (4.02,16.00) (4.03,16.00) (4.04,16.00) (4.05,16.00) (4.06,16.00) (4.07,16.00)
  (4.08,15.98) (4.09,16.01) (4.10,16.00) (4.11,16.00) (4.12,16.00) (4.13,16.00)
  (4.14,16.00) (4.15,16.00) (4.16,15.99) (4.17,16.00) (4.18,15.99) (4.19,16.01)
  (4.20,11.84) (4.21,1.02) (4.22,1.00) (4.23,1.00) (4.24,1.00) (4.25,1.00)
  (4.26,1.00) (4.27,1.00) (4.28,1.00) (4.29,1.00) (4.30,1.00) (4.31,1.00)
  (4.32,1.00) (4.33,0.99) (4.34,1.00) (4.35,1.00) (4.36,1.11) (4.37,16.00)
  (4.38,15.98) (4.39,16.01) (4.40,16.00) (4.41,13.95) (4.42,16.07) (4.43,3.91)
  (4.44,1.47) (4.45,1.00) (4.46,1.00) (4.47,0.94) (4.48,0.95) (4.49,0.94)
  (4.50,0.42)
  (4.51,0.42)
};
\addlegendentry{\lld}
\end{axis}
\end{tikzpicture}
\caption{CPU core utilization over time for \mold and \lld on
  the Firefox debug build. Each step is the average number of active
  CPU cores over a 10\,ms interval, computed from changes in the
  \code{usage\_usec} counter of a dedicated cgroup.}
\Description{A step chart of active CPU cores over elapsed time. \mold
keeps as many as 32 cores active for much of its approximately
0.9-second link. lld remains near one active core for most of its
approximately 4.5-second link, interrupted by several brief bursts of
about 16 active cores.}
\label{fig:cpu-util}
\end{figure}

Figure~\ref{fig:cpu-util} visualizes the difference in
CPU utilization over time: \mold keeps many cores active
for most of its 0.9\,s execution, while \lld runs predominantly
on a single core with brief multi-core bursts, spread over
4.5\,s. This is Amdahl's law in action: \lld's serial portions
dominate its execution time regardless of core count, while
\mold's pervasive data parallelism keeps all worker threads
busy.

\begin{table}[t]
\centering
\caption{Cumulative CPU time (user + system) and wall-clock time
  on the Firefox debug build (seconds).}
\label{tab:cpu-time}
\small
\begin{tabular}{@{}rrrrr@{}}
\toprule
& \multicolumn{2}{c}{CPU time} & \multicolumn{2}{c}{Wall clock} \\
\cmidrule(lr){2-3} \cmidrule(l){4-5}
Threads & \mold & \lld & \mold & \lld \\
\midrule
1  & 12.2 & 11.4 & 12.2 & 11.4 \\
8  & 14.3 & 13.4 & 1.9  & 4.8  \\
16 & 16.0 & 14.7 & 1.2  & 4.4  \\
32 & 21.1 & 18.1 & 0.9  & 4.4  \\
64 & 38.2 & 25.5 & 0.9  & 4.5  \\
\bottomrule
\end{tabular}
\end{table}

\revised{Table~\ref{tab:cpu-time} shows\revpoint{A-W4} the CPU cost of parallel speedup.
At 32 threads, \mold achieves a 13.5x latency reduction by using 73\%
more CPU time than the one-thread run. At 64 threads, CPU time grows a
further 80\%, from 21.1\,s to 38.2\,s, with wall-clock time unchanged
at 0.9\,s. Adding threads beyond 32 increases CPU time without making
the linker faster.}

\revised{Hardware performance counters\revpoint{A-W5}, collected
with Linux \code{perf} and AMD instruction-based sampling
(IBS), attribute the
plateau to two compounding effects. On the demand side, more
workers traverse \mold's metadata at once, so the combined working
set grows while cache capacity does not: from 32 to 64 threads,
\mold retires only 9\% more instructions but incurs 37\% more
demand loads served from DRAM. On the supply side, the DRAM fill
rate rises by only 38\%, even though twice as many threads now
wait on memory. For small random reads, DRAM throughput can saturate
well before the data bus is fully utilized, and the 32-thread run
already appears close to this limit: its sampled DRAM loads average 570
cycles (about 180\,ns), roughly 40\% higher than the machine's unloaded
random-access latency of about 130\,ns. The additional requests at 64
threads therefore mostly increase waiting time rather than DRAM
throughput: the mean
latency inflates 1.9x, to about 1{,}070 cycles. The longer stalls
reduce IPC from 0.85 to 0.53, explaining the extra CPU time.
In sum, the larger working set causes 37\% more demand loads to reach
DRAM, so the 38\% increase in DRAM fill rate is almost entirely
absorbed by the additional traffic, leaving wall-clock time unchanged.}

\subsection{Ablation Study}
\label{sec:eval-ablation}

To quantify the contribution of each parallelized pass, we measure
\mold's performance when individual passes are forced to run
single-threaded while the rest remain parallel. For each pass,
Table~\ref{tab:ablation} shows the time spent in the pass itself,
run in parallel and serialized, and the effect on end-to-end link
time, for the Firefox debug workload.

\begin{table}[t]
\centering
\caption{Ablation study on the Firefox debug build. Each row shows the
  time spent in the named pass when it runs in the default parallel
  configuration (``Par.'') and when it is forced to run
  single-threaded while all other passes remain parallel (``Ser.''),
  along with the resulting end-to-end link time, all in seconds.
  ``Overhead'' is the increase in link time relative to
  fully parallel \mold.}
\label{tab:ablation}
\small
\begin{tabular}{@{}lrrrr@{}}
\toprule
Serialized pass & Par. & Ser. & Total & Overhead \\
\midrule
None (fully parallel)       & ---  & ---   & 0.91  & ---      \\
Input file parsing          & 0.06 & 0.54  & 1.38  & $+$52\%  \\
Symbol resolution           & 0.11 & 1.26  & 2.05  & $+$125\% \\
String merging              & 0.03 & 0.42  & 1.27  & $+$40\%  \\
Section garbage collection  & 0.06 & 1.09  & 1.95  & $+$114\% \\
Relocation scanning         & 0.10 & 0.50  & 1.30  & $+$43\%  \\
Output copy + reloc.\ apply & 0.24 & 5.18  & 5.84  & $+$542\% \\
Build-ID computation        & 0.02 & 0.58  & 1.47  & $+$62\%  \\
\bottomrule
\end{tabular}
\end{table}

Every pass parallelizes, with pass-level speedups ranging from 5x
to 29x at the default 32 threads. The end-to-end impact of serializing
a pass is determined mostly by the pass's share of total work.
At one extreme, output copy adds
$+$542\% to the end-to-end link time when serialized. At
the other, build-ID computation scales the best of all the passes
(29x) yet costs only $+$62\% when serialized, because it is a tiny
fraction of the link. The
cumulative effect of parallelizing \emph{all} passes is what
produces the large overall speedup: no single pass explains it, and
even passes with modest individual overhead contribute to the
end-to-end link time when serialized.

\subsection{Where Does lld Spend Its Time?}
\label{sec:eval-lld}

To understand why \lld takes 4.9x as long as \mold, we profile
both linkers on the Firefox debug workload using their built-in
per-pass instrumentation. Table~\ref{tab:lld-breakdown} shows the
results. Since \lld resolves symbols inside its file-parsing
loop, its profile cannot report symbol resolution separately; the
first row therefore combines parsing and symbol resolution.

\begin{table}[t]
\centering
\caption{Per-pass time breakdown comparison between \mold and \lld
  on the Firefox debug build (wall-clock seconds).}
\label{tab:lld-breakdown}
\small
\begin{tabular}{@{}lrrr@{}}
\toprule
Pass & \mold & \lld & Speedup \\
\midrule
Parsing + symbol resolution     & 0.17 & 1.29 & 7.6x \\
String merging                  & 0.03 & 0.31 & 10.3x \\
Symbol version processing       & 0.00 & 0.24 & --- \\
Section garbage collection      & 0.06 & 0.85 & 14.2x \\
Relocation scanning             & 0.10 & 0.10 & 1.0x \\
Output copy + reloc.\ apply     & 0.23 & 0.50 & 2.2x \\
Other                           & 0.31 & 1.14 & 3.7x \\
\midrule
\textbf{Total} & \textbf{0.90} & \textbf{4.43} & \textbf{4.9x} \\
\bottomrule
\end{tabular}
\end{table}

The 3.53\,s difference is distributed across the pipeline: no
single category accounts for more than a third of it. Parsing
and symbol resolution make the largest absolute contribution. \lld spends 1.29\,s, or 29\% of its total time, in
its largely sequential parse-and-resolve loop, versus 0.17\,s for
\mold. This loop appears in Figure~\ref{fig:cpu-util} as the
single-core interval during \lld's first 1.3\,s. Among passes
with measurable times in both linkers, section garbage collection
has the largest relative gap, at 14.2x. \lld's mark phase is
single-threaded and takes 0.85\,s, whereas \mold marks and sweeps
in parallel and completes the pass in 0.06\,s. Symbol version
processing is also serial in \lld and costs 0.24\,s; \mold
performs it in parallel, and its time rounds to 0.00\,s at the
precision shown.

Passes that \lld also parallelizes show a mixed picture.
Relocation scanning takes 0.10\,s in both linkers, showing that
\mold does not have a uniform constant-factor advantage.
String merging shows the widest of these gaps, at 10.3x:
\lld parallelizes only part of
the pass and, even at one thread, takes 1.7\,s versus \mold's
0.4\,s. The output-copy and relocation-application pass is 2.2x
faster in \mold. \lld parallelizes across output sections but not within
them, so each section is processed by a single thread regardless
of how much work it requires. This coarse granularity
follows from \lld's parallel-for implementation: a parallel-for
invoked from within another parallel-for executes serially. This
leaves the computationally intensive \code{.eh\_frame\_hdr}
section to a single thread, creating a serial tail. \mold's
parallel runtime supports nested parallel-for loops, allowing it
to parallelize both across and within output sections.

The remaining 26\% of \lld's time, 1.14\,s, is spread across
smaller, mostly sequential steps such as section layout,
local symbol processing, and address-dependent content
finalization; the corresponding remainder in \mold is 0.31\,s.
Overall, \lld takes 4.9x as long as \mold not because of one
dominant bottleneck, but because serial or incompletely parallel
work appears throughout the pipeline.

\subsection{Range Extension Thunks}
\label{sec:eval-thunks}

Range extension thunks are created during section layout
computation, making it difficult to isolate their overhead as a
separate pass in a profiler. We instead disabled thunk creation in
both \mold and \lld and compared link times with and without it.
The resulting executables are broken, since out-of-range branches
no longer reach their targets, but only the link time matters
here. We measure the
ARM64 Firefox debug link, cross-linked on the x86-64 machine.

Table~\ref{tab:thunks} shows the results. Thunks cost \mold
0.07\,s but \lld 1.01\,s, a 14x
difference. To determine how much of the gap comes from
the algorithm itself and how much from its parallelization, we
repeated the measurement with both linkers restricted to one
thread. Run single-threaded, \mold's
thunk machinery costs 0.51\,s versus \lld's 0.92\,s, so \mold's
linear scan algorithm with a cleanup pass beats \lld's repeated rescans by
1.8x on algorithm alone (\S\ref{sec:thunks}). Parallelism accounts for the
rest: \mold scans each batch's relocations in parallel, cutting its
overhead 7x to 0.07\,s, whereas \lld's scan is single-threaded,
leaving its overhead essentially unchanged regardless of thread
count.

\begin{table}[t]
\centering
\caption{Range extension thunk overhead on the ARM64 Firefox
  debug build, measured by re-linking the same input with thunk
  creation disabled.}
\label{tab:thunks}
\small
\begin{tabular}{@{}lrrr@{}}
\toprule
Linker & Normal & No thunks & Overhead \\
\midrule
\mold & 1.05\,s & 0.98\,s & $+$0.07\,s \\
\lld  & 5.56\,s & 4.55\,s & $+$1.01\,s \\
\bottomrule
\end{tabular}
\end{table}

\section{Compatibility with GNU ld}
\label{sec:eval-compat}

As a compatibility test, we attempted to build all
19{,}422 packages in the Gentoo Linux package repository inside a
container whose system linker was replaced with \mold; 74 of them
\revised{failed either to build or to pass their own test
suite\revpoint{B-Q1}} with \mold but not with \gnuld.

Of the 74 failures, 25 involved bootloaders, firmware, kernels,
and other non-userland software, which is currently outside
\mold's scope (\S\ref{sec:limitations}). Nine of the remaining
49 are false positives: the binaries work, but their test
suites inspect the generated ELF files and reject harmless
differences. The other 40 stem from a long tail
of causes, including unsupported options such as \code{-b binary}
and differences in library search behavior. Only two packages
failed because \mold's parallel symbol resolution
(\S\ref{sec:symres}) selects different archive members than
\gnuld does.

\section{Limitations}
\label{sec:limitations}

\revised{\mold implements\revpoint{C-Q1} a subset of the GNU linker script language that is
sufficient for almost all userland programs. It does not yet
implement \code{SECTIONS}-based layout control or the
\code{MEMORY} and \code{OVERLAY} commands, so \mold cannot be used where
explicit layout control is required, notably for Linux kernel
development and many embedded systems.}

\revised{\mold currently\revpoint{C-Q2} supports only ELF, but little in its design is
ELF-specific: every pass operates on sections, symbols, relocations,
and archives, structures that other object formats such as COFF (Windows) and
Mach-O (macOS) share. Indeed, an earlier version of \mold included a Mach-O
port built on the same parallel pass structure. We discontinued the
port after Apple replaced its system linker with a new one that is
substantially faster~\cite{ldprime}, which reduced the demand
for a third-party fast linker on macOS. We expect the techniques
described in this paper to carry over to both COFF and
\mbox{Mach-O}.}

By default, \mold limits itself to 32 threads, as we
observe diminishing returns beyond this point
(Figure~\ref{fig:scalability}).
Investigating performance on high-core-count server processors,
whose additional memory channels should raise the random-access
ceiling observed in \S\ref{sec:eval-scale}, is future work.

\section{Related Work}
\label{sec:related}

The most closely related systems are the three major open-source ELF
linkers, \gnuld, \gold, and \lld, described in
\S\ref{sec:background}; Table~\ref{tab:parallel-comparison}
compares which of their passes are parallelized.
Song~\cite{maskray2021lld} analyzed \lld's performance
characteristics and identified several phases that remain
sequential, all of which \mold parallelizes.

Wild~\cite{wild} is a recent ELF linker written in Rust that also
targets parallel linking performance. As of this writing (mid 2026),
Wild is under
active development and still lacks some features available in
production linkers.

Levine's \emph{Linkers and Loaders}~\cite{levine2000linkers} is still
the standard introduction to how linkers work, though parts of it no
longer reflect current practice a quarter century after its
publication.

\revpoint{A-W3}\revised{Incremental linking takes a different route to fast turnaround. It
patches the previous output in time proportional to the size of a
change~\cite{quong1991inclink}; \gold supports it, and Wild's
long-term goals include it. \mold instead focuses on full-link speed.
Incremental linking adds bookkeeping overhead to every link, and many
common changes still force a full relink. For example, \gold falls
back to a full link when a function definition is removed or
preallocated patch space runs out. A fast full link therefore remains
important even for an incremental linker.}

\revised{Split DWARF reduces linker work in another way. Instead of storing
all debug information in object files, the compiler writes most of
it to separate \code{.dwo} files, which the debugger reads
directly and the linker never touches. The scheme was introduced
in part to address long link times caused by large amounts of
debugging information~\cite{coutant2014splitdwarf}. Adoption
remains limited, as it is disabled by default and the \code{.dwo}
files complicate build and debugging workflows. The approach is
complementary to making the linker itself faster.}

\revised{Another common workaround for long link times is to split a
program into shared libraries, so that most changes relink one
small library rather than the full executable. A documented
example is Chromium's component
build~\cite{chromecomponentbuild}. The split comes at the cost of larger
binaries, slower loading, and added build and code
complexity. By making multi-gigabyte
full links complete in a few seconds, \mold makes such
link-time-driven decomposition less compelling.}

Tallam et al.~\cite{tallam2010safeicf} describe the ICF
implementation in the gold linker. \mold uses a different,
parallel algorithm (\S\ref{sec:icf}).

ThinLTO~\cite{johnson2017thinlto} addresses the scalability of
link-time optimization through partitioned compilation guided by
module summaries. \mold supports ThinLTO and full LTO as a black-box
pass, orthogonal to its linking parallelism.

BOLT~\cite{panchenko2019bolt} is a post-link binary optimizer that
reorganizes code layout based on profile data. BOLT operates on
linker output and is complementary to \mold.

Build systems such as Bazel~\cite{bazel}, Buck~\cite{buck}, and
Ninja~\cite{ninja} exploit parallelism at the level of compilation
tasks. \mold is complementary: it
accelerates the linking step that typically cannot be parallelized at
the build-system level because it is a single invocation.

\section{Conclusion}
\label{sec:conclusion}

We have presented \mold, a production Unix/Linux linker that
structures every major pass as a parallel-for loop over homogeneous
data. On real-world workloads, \mold is 2.4--16.1x faster than the
state-of-the-art \lld, even with \mold's portable
system-level optimizations applied to \lld, and up to 112x faster
than the traditional \gnuld. This gap has no single cause.
Prior linkers leave serial work in many different passes, and fully
parallelizing these passes would require both new algorithms and substantial
architectural changes. By building a new linker around pervasive
parallelism, we show how every major linker pass can be parallelized
and that doing so yields substantial end-to-end speedups.

\begin{acks}
We thank the anonymous reviewers for their feedback on this paper.
We also thank the many contributors to \mold, and the project's
GitHub sponsors, in particular Cybozu, \mbox{G-Research}, Mercury, Signal
Slot, Ahrefs, Jinkyu Yi, Wei Wu, and Pedro Navarro, for their
support. Claude Code was used to help draft and edit parts of this
paper and to develop the benchmarking scripts for the evaluation.
\end{acks}

\bibliographystyle{ACM-Reference-Format}
\bibliography{references}

\makeatletter
\if@ACM@anonymous\else
  
\appendix
\section{Artifact Appendix}
\label{sec:artifact}

\subsection{Abstract}

The artifact contains a container image with the evaluated linker
configurations, captured inputs for the nine workloads in
Table~\ref{tab:workloads}, and scripts
reproducing Tables~\ref{tab:e2e}, \ref{tab:rss}, \ref{tab:cpu-time},
\ref{tab:ablation}, and~\ref{tab:thunks} and Figures~\ref{fig:scalability}
and~\ref{fig:cpu-util}. It runs on x86-64 Linux and emits results in CSV
format. Exact setup and execution commands are provided in the
accompanying README.

\subsection{Artifact check-list (meta-information)}

{\small
\begin{itemize}
  \item {\bf Algorithm: } Data-parallel ELF linking
  \item {\bf Program: } Nine workloads listed in
    Table~\ref{tab:workloads}; included with the artifact
  \item {\bf Binary: } x86-64 Linux binaries for \mold 2.42.0,
    \lld 22.1.8, and \gnuld and \gold 2.46.1; included in the
    container image
  \item {\bf Data set: } Captured linker inputs for nine workloads
  \item {\bf Run-time environment: } x86-64 Linux; Podman or Docker
  \item {\bf Hardware: } At least 64\,GiB RAM and 200\,GiB free disk;
    128\,GiB RAM and NVMe storage recommended
  \item {\bf Run-time state: } Otherwise-idle host;
    transparent huge pages is not disabled
  \item {\bf Execution: } Shell and Python scripts; one warmup
    followed by five measured runs
  \item {\bf Metrics: } Wall-clock and CPU time, peak RSS, and CPU
    utilization
  \item {\bf Experiments: } Tables~\ref{tab:workloads}, \ref{tab:e2e}, \ref{tab:rss}, \ref{tab:cpu-time},
     \ref{tab:ablation}, and~\ref{tab:thunks}; Figures~\ref{fig:scalability}
     and~\ref{fig:cpu-util}
  \item {\bf How much time is needed to prepare workflow
    (approximately)?: } Network-dependent download and extraction;
    add about 30 minutes if rebuilding the container image
  \item {\bf How much time is needed to complete experiments
    (approximately)?: } About 4 hours for the core results and
    45 minutes for the supplemental results
  \item {\bf Publicly available?: } Yes
  \item {\bf Code licenses: } MIT,
    Apache-2.0 with LLVM exception, and GPL-3.0-or-later
  \item {\bf Data licenses: } Upstream
    workload licenses; included with each workload
  \item {\bf Archived (provide DOI)?: }
    \href{https://doi.org/10.5281/zenodo.21882261}{doi:10.5281/zenodo.21882261}
\end{itemize}
}

\subsection{Description}

The artifact is archived on Zenodo at
\url{https://doi.org/10.5281/zenodo.21882261}. Running it requires
an x86-64 Linux machine with Podman or Docker, at least 64\,GiB of
RAM, and approximately 200\,GiB of free space on an ext4 filesystem;
128\,GiB of RAM and NVMe storage are recommended. All linker binaries
and utilities are included in the container image, and the
Containerfile records their provenance. Exact workload versions,
licenses, and build configurations are documented in the README.

\subsection{Installation}

Unpack the artifact archive onto an ext4 filesystem and load the
container image:

{\small
\begin{verbatim}
  tar xf bench.tar.gz
  cd bench
  podman load -i mold-ae.tar
\end{verbatim}
}

As a basic test, link the smallest workload once with every
linker, from the unpacked benchmark directory (about a minute):

{\small
\begin{verbatim}
  podman run --rm -v $PWD:/bench \
    --security-opt label=disable \
    -e PROJECTS=godot-release -e RUNS=1 \
    mold-ae with-thp ./tab4-bench.sh
\end{verbatim}
}

\noindent
It prints one CSV row per linker; a failed link would appear as an
empty time field. With \code{RUNS=1}, the sole value is the warmup
run and is intended only as a quick compatibility check.

\subsection{Workflow and expected results}

The README maps each paper result to the script that reproduces it and
gives the exact container command. Timing experiments use one warmup
run followed by five measured runs, whose median corresponds to the
reported result. Absolute times depend on the machine, but \mold
should outperform \lld on all workloads, continue scaling after \lld
plateaus, and slow down when any pass listed in
Table~\ref{tab:ablation} is serialized.

\subsection{Notes}

The benchmark inputs must reside on a bind-mounted ext4 filesystem.
Storing them in the container overlayfs disables huge-page-backed file
mappings and introduces I/O overhead that would skew the comparison.

\fi
\makeatother

\end{document}